\documentclass[a4paper,11pt]{article}
\pdfoutput=1 

\usepackage{jheppub} 

\usepackage[T1]{fontenc} 
\usepackage{multirow}
\usepackage{booktabs}

\title{\boldmath Multi-Higgs-Doublet Models and\\ Singular Alignment}

\author{Werner Rodejohann}
\author{and Ulises Salda\~na-Salazar}
\affiliation{ 
Max-Planck-Institut f\"ur Kernphysik,\\
Postfach 103980, D-69029 Heidelberg, Germany
}

\emailAdd{werner.rodejohann@mpi-hd.mpg.de}
\emailAdd{ulises.saldana@mpi-hd.mpg.de}

\abstract{
We consider a 4-Higgs-doublet model in which each Higgs doublet gives mass to one of the 
fermion sets 
$\{ m_t\}$, $\{ m_b,m_\tau,m_c \}$, $\{ m_\mu,m_s \}$, and $\{ m_d,m_u,m_e \}$. The sets have the feature that within each of them the masses are similar. Our model explains the mass hierarchies of the sets 
by hierarchies of the vacuum expectation values of the Higgs doublets associated to them. All Yukawa  couplings are therefore of order one. 
 Neutrino masses are generated by a type-I seesaw mechanism with PeV-scale singlet neutrinos.  
To avoid the appearance of tree-level flavour changing neutral currents, we assume that all Yukawa matrices are singularly aligned in flavour space. We mean by this that the Yukawa matrices are given as linear combinations of the rank $1$ matrices that appear in the singular value decomposition of the mass matrix. In general, singular alignment allows to avoid flavour changing neutral currents in models with multiple Higgs doublets. 
}

\begin{document} 
\maketitle
\flushbottom

\section{Introduction}
\noindent
An understanding of fermion masses and mixing is still lacking. In particular, the mass values display unexplained patterns and hierarchies; this is the case when one considers the three generations as well  as the species\footnote{That is, any of the four masses within the same generation.}: 

\begin{center}
\begin{tabular}{ccccccc}
  & \multicolumn{6}{c}{$\quad\xrightarrow{\quad \text{intergeneration}\; \quad }$} \\
 $v_\text{EW}$ & $\sim$ & $m_t$ & $\gg$ & $m_c$ & $\gg$ & $m_u$ \\
  & \multirow{6}{*}{\rotatebox[origin=c]{-90}{$\xrightarrow[\quad \text{interspecies} \quad]{}$}}   
  & \rotatebox[origin=c]{-90}{$\gg$} & &  \rotatebox[origin=c]{-90}{$\gg$} & &  \rotatebox[origin=c]{-90}{$<$} \\
  &  & $m_b$ & $\gg$ & $m_s$ & $\gg$ & $m_d$ \\
  &    & \rotatebox[origin=c]{-90}{$>$} & &  \rotatebox[origin=c]{-90}{$<$} & &  \rotatebox[origin=c]{90}{$<$} \\
  &  & $m_\tau$ & $\gg$ & $m_\mu$ & $\gg$ & $m_e$ \\ 
  &    & \rotatebox[origin=c]{-90}{$\gg$} & &  \rotatebox[origin=c]{-90}{$\gg$} & &  \rotatebox[origin=c]{90}{$\ll$} \\
  &  & $m_{\nu 3(2)}$ & ? & $m_{\nu 2(1)}$ & ? & $m_{\nu 1(3)}$ \\ 
\end{tabular} 
\end{center}

	We can summarize the situation by asking the following questions:
\begin{itemize}
	\item Why is the top quark mass the \textit{only fermion} mass of the order 
	of the electroweak (EW) scale, $m_t \approx {v_\text{EW}}$ with $v_\text{EW} \simeq 174 \text{ GeV}$?
	\item Why is the top quark mass so much heavier than the rest of fermion masses, $m_t \gg m_f $?
	\item Why do all \textit{charged} fermions satisfy the hierarchy, $m_3 \gg m_2 \gg m_1$?
	\item Why have the down-type quarks and charged leptons similar masses, $m_{d_i} \sim m_{e_i}$ ($d_{1,2,3}= d, s, b$, $e_{1,2,3} = e, \mu, \tau$)?
	\item Why are for the first generation the masses (except for neutrinos) closer to each other than for the other two generations, $m_d \sim m_u \sim m_e$ versus $m_c \gg m_s \sim m_\mu$ and $m_t \gg m_b \sim m_\tau$?
	\item What could the \textit{interspecies}~hierarchy, e.g.\ $m_t \gg m_b > m_\tau \gg m_{\nu 3}$, be telling us?
	\item Why are neutrino masses much smaller than the charged fermions\footnote{Once the neutrino mass ordering and hierarchy is determined, it is likely that additional questions will arise.}, $m_\nu \sim 10^{-7} m_e $? 
\end{itemize}
This is commonly referred to as the problem of mass~\cite{Weinberg:1977hb}. 

Part of the mystery lies in the contrast of expecting Yukawa couplings to be order one, $y_f = {\cal O}(1)$, whereas the observed values with a 
single Higgs doublet are much smaller than $1$, except for the top quark, $y_f \ll 1$ ($f\neq t$). 
In the following, we assume Yukawa couplings to be order one, $y_f = {\cal O}(1)$, and try to  understand the fermion mass patterns through a theory with multiple Higgs doublets. 
The most extreme approach along this line would be the "private Higgs" scenario, in which among other things, for each fermion a Higgs doublet is introduced \cite{Porto:2007ed,Porto:2008hb}, see also \cite{Camargo-Molina:2017klw,Diaz-Cruz:2019emo,Hill}. The mass hierarchies are explained by hierarchies of vacuum expectation values of the individual Higgs doublets: $m_f \simeq v_f$, where $v_f$ is the vacuum expectation value of the Higgs that is responsible for the fermion $f = u,d,c,s,t,b,e,\mu,\tau$.

In general, in a model with $N$ Higgs doublets, $\Phi_i$ ($i=1,2,\ldots,N$), where each of their neutral 
	components acquires a vacuum expectation value (vev), $\langle \Phi^0_j \rangle =v_j \,e^{i\theta_j}$, 
	a relation among these vacua is satisfied: 
	\begin{align}
		\sum_{i=1}^N v_i^2 = v^2_\text{EW} \;.
	\end{align}
	Here $v_\text{EW} \simeq 174 \text{ GeV}$, $v_i \geq 0$, and all doublets share the same hypercharge $Y=\tfrac{1}{2}$.
	
	Now, if we consider that 
	each single Higgs is fully responsible for the mass of one single fermion (where $N$ should
	equal the number of fermions in the theory), then the previous relation is modified to 
	\begin{align}
		\sum_{i=1}^N \frac{m_i^2}{y_i^2} = {v_\text{EW}^2} \;.
	\end{align}		
	Furthermore, if we consider that Yukawa couplings should be order one numbers, $y_f = {\cal O}(1)$, we could
	approximately say that, to good approximation
	\begin{align}\label{eq:mvr}
		\sum_{i=1}^N {m_i^2} \approx {v_\text{EW}^2} \;.
	\end{align}	
	In the case of the Standard Model (SM), with $N=12$ fermions, the previous equation is fulfilled.
	We will call this relation the mass-vacuum relation. 
	An amusing possibility from this relation is that 
	if all $N$ doublets have the same vev, 
	one would have $N$ fermions with mass of about $174/\sqrt{N}$ GeV, which would be about 50 GeV for 12 fermions. If two doublets have vev $v_\text{EW}/\sqrt{2}$ and the rest a vanishing vev, then there would be two fermions with mass $v_\text{EW}/\sqrt{2} \simeq 123$ GeV. In turn, if only one doublet has a vev, there is only one fermion with mass $v_\text{EW}$. Forcing the mass-vacuum relation to be fulfilled and assuming that only one Higgs acquires a vev leaves hardly any mass for the other fermions and explains the top quark's dominance. Moreover, this same argument could help us to understand why neutrinos are so light when assumed as Dirac fermions\footnote{We will not focus too much on neutrino masses in this paper, there are several possibilities to incorporate them in multi-Higgs-doublet models, see e.g.\ \cite{Wang:2006jy,Campos:2017dgc,Hill}.}.

The particle content in the main scenario discussed in this paper is smaller than that for a private Higgs-like scenario. 
Our observation is that the fermion masses can be grouped into  four different sets: 
$\{ m_t\}$, $\{ m_b,m_\tau,m_c \}$, $\{ m_\mu,m_s \}$, and $\{ m_d,m_u,m_e \}$. In each set the masses are quite similar and can in fact be explained by similar ${\cal O}(1)$ Yukawa couplings to 
an individual Higgs doublet $\Phi_{t,b,\mu,d}$. Such a 4-Higgs-Doublet Model has to the best of our knowledge not been considered before. We find several attractive and testable features of the model, and demonstrate that it is not in conflict with measured Higgs couplings and other tests. Our model traces the hierarchy of the mass values of the different fermion sets to hierarchies of vevs  of their respective Higgs doublets. 
We show that the smaller vevs can be induced by the larger vevs, and the hierarchy among them arises because the 
four vevs are protected by different symmetries. 

The main problem in multi-Higgs doublet models is of course the presence of flavour changing neutral currents (FCNC). Theories which through the use of symmetries naturally avoid those FCNC are said to possess Natural Flavour Conservation (NFC). 
Options to evade FCNC include, next to arranging the additional scalar particles to be very heavy,  
suppressing dangerous Yukawa couplings \cite{Cheng:1987rs}, 
separating the Yukawa matrices such that only one scalar 
doublet couples to a given right-handed fermion field \cite{Paschos:1976ay,Glashow:1976nt}, or Yukawa alignment~\cite{Pich:2009sp,Penuelas:2017ikk}, in which the different Yukawa matrices are 
proportional to each other. 
As a proof of principle that FCNC can be entirely avoided in our setup, 
we assume here another solution. We note that if the Yukawa matrices are proportional to 
any of the rank-one matrices that appear in the singular value decomposition of the fermion mass matrices, FCNC are absent. We denote this as "singular alignment". 

The paper is organized as follows: In Section \ref{sec:SA} we  present  singular alignment and discuss some of its features. The model with four Higgs doublets to explain the masses of the individual sets 
$\{ m_t\}$, $\{ m_b,m_\tau,m_c \}$, $\{ m_\mu,m_s \}$, and $\{ m_d,m_u,m_e \}$ is presented and analyzed in Sec.\ \ref{sec:model}. Conclusions are presented in Sec.\ \ref{sec:concl}, and some technical details are delegated to Appendices.

\section{\label{sec:SA}Singular Alignment}
 \noindent 
In general, having multiple Higgs doublets coupling to fermions with the same electric charge will produce tree-level FCNC, which are experimentally strongly constrained. 
Three main possibilities to overcome this problem have typically been studied: 
(i) assume "dangerous" Yukawa couplings to be sufficiently 
suppressed at tree-level~\cite{Cheng:1987rs};   
(ii) assume the corresponding Yukawa matrices of each type of fermion (up-type quarks, down-type quarks and charged leptons) to be 
proportional to the mass matrix~\cite{Pich:2009sp,Penuelas:2017ikk}; 
(iii) impose an adequate symmetry such that each fermion type couples exactly to one of the doublets 
\cite{Paschos:1976ay,Glashow:1976nt}. 
In the following, we comment only on the last two possibilities and introduce singular alignment. 

Let us start from the most general case for a Yukawa
Lagrangian in a $N$HDM,
\begin{align}
	-{\cal L}_Y \supset \sum_{a=1}^N \overline{F}_L {\bf Y}^f_a f_R \Phi_a + \text{  H.c.},
\end{align}
where $F_L$ and $f_R$ are three dimensional vectors in family space and transform as
a doublet and as a singlet under $SU(2)_L$, respectively.
The $N$ Higgs doublets acquire a vev, $v_a = \langle \Phi^0_a \rangle$. 
In general, Yukawa couplings will couple all fermions to all  Higgses. 
Therefore, the most general form of a mass matrix is 
\begin{align} \label{eq:MassMatrixNHDM}
	{\bf M} = v_1 {\bf Y}_1 + v_2 {\bf Y}_2 + \cdots + v_N {\bf Y}_N \; .
\end{align} 
Each Yukawa matrix, ${\bf Y}_i$, is a $3 \times 3$, arbitrary, and complex matrix with
rank 3. The appearance of tree-level FCNC is automatic within this setup as diagonalization
of the mass matrices does not mean, in general, simultaneous diagonalization
of the individual Yukawa matrices. However, to avoid introducing dangerous tree-level
FCNC the following can be done:

	\paragraph{NFC theories:} Adequate symmetries are imposed in such a way that
	each of the three charged fermions will only couple to a single Higgs~\cite{Paschos:1976ay, Glashow:1976nt}, i.e.\ for each fermion type holds
	\begin{align}
	{\bf M} =  v_k {\bf Y}_k \; ,
	\end{align} 
	where no sum over $k$ is intended.
	In this case diagonalization of the l.h.s.\ means diagonalization of the r.h.s. For $N$ Higgs doublets,
	the easiest way to achieve this is via a symmetry of the form 
	\begin{align}
		Z^{(1)}_2 \times Z^{(2)}_2 \times \cdots \times Z_2^{(\ell)} \; ,
	\end{align}
	where in order for this symmetry
	to be \textit{realizable} $\ell = N-1$ should hold.
	Realizable symmetries are a set of allowed discrete symmetries
	of the scalar potential which have no accidental 
	larger groups that could give rise, for example, to massless Goldstone bosons~\cite{Ivanov:2011ae}.

	Now, before turning to the next possiblity, let us comment on the 
	Singular Value Decomposition (SVD) of a mass matrix: 
	\begin{align}
		{\bf M} = {\bf L}^\dagger {\bf \Sigma} {\bf R} \; .
	\end{align}
	Here ${\bf L}$ and ${\bf R}$ are unitary matrices which rotate independently
	the left- and right-handed fermion fields and ${\bf \Sigma} = \text{diag} (m_1 ,m_2,m_3)$
	with $m_i > 0$.
	Realize that the SVD may also be written as a sum of three rank $1$ matrices,
	\begin{align}
		{\bf M} = \sum_i m_i {\bf L}^\dagger {\bf P}_i {\bf R} \; ,
	\end{align}
	where ${\bf P}_i$ are three projector operators, ${\bf P}_i^2 = {\bf P}_i$ and $\sum_i {\bf P}_i = 1_{3\times 3}$, which have the form
	\begin{align}
		{\bf P}_1 = \begin{pmatrix}
			1 & 0 & 0 \\
			0 & 0 & 0\\
			0 & 0 & 0
		\end{pmatrix} \; ,\qquad
		{\bf P}_2 = \begin{pmatrix}
			0 & 0 & 0 \\
			0 & 1 & 0\\
			0 & 0 & 0
		\end{pmatrix} \; , \qquad
		{\bf P}_3  = \begin{pmatrix}
			0 & 0 & 0 \\
			0 & 0 & 0\\
			0 & 0 & 1
		\end{pmatrix} .
	\end{align}
	In the following, we will denote each rank $1$ matrix appearing in the SVD by 
	\begin{align}\label{eq:11}
		{\bf \Delta}_i = {\bf L}^\dagger {\bf P}_i {\bf R} \;,
	\end{align}
	 and call it singular matrix.
	
	\paragraph{Yukawa Alignment:}
	As each Yukawa term in Eq.~\eqref{eq:MassMatrixNHDM} is a rank $3$ matrix, 
	a second possibility to avoid FCNC, is to assume that each of them is proportional to 
	the full SVD \cite{Pich:2009sp,Penuelas:2017ikk}: 
	\begin{align} \label{eq:NormalYukAlig}
		 {\bf Y}_i = \frac{\zeta_i}{v_i} \left(  m_1 {\bf \Delta}_1 + m_2  {\bf \Delta}_2 + m_3  {\bf \Delta}_3 \right)  .
	\end{align}
	Here the $\zeta_i$ are real and we have that 
\begin{align} 
	{\bf M} = \left(\sum_{j=1}^N {\zeta}_j \right) \left( m_1 {\bf \Delta}_1 + m_2 {\bf \Delta}_2 + m_3{\bf \Delta}_3 \right) .
\end{align} 
	Diagonalization of the l.h.s.\ means diagonalization of the r.h.s. This is understandable
	as each Yukawa matrix is rank $3$ and thus if related to the singular matrices should be 
	composed of the three independent singular matrices. Furthermore, one has the constraint 
	\begin{align}
		\sum_{j=1}^N {\zeta}_j  = 1 \; .
	\end{align}
	
	\paragraph{Singular Alignment:}	
	A more general scenario  is that in which each Yukawa matrix is given by a linear combination of
	the singular matrices, i.e.\ 
	\begin{align} \label{eq:SingAlig}
		 {\bf Y}_i = \left( \eta_i  {\bf \Delta}_1 + \Omega_i {\bf \Delta}_2 +\Lambda_i {\bf \Delta}_3 \right)  .
	\end{align}
	Appendix \ref{sec:appA} gives a straightforward  proof of the absence of FCNC in case the Yukawa matrices take this form. 
	Comparing with the full mass matrix, which can be written as 
	\begin{align}
		{\bf M} = m_1 {\bf \Delta}_1 + m_2 {\bf \Delta}_2 + m_3 {\bf \Delta}_3 \; ,
	\end{align}
	we identify 
	\begin{equation} \label{eq:massesNHDM}
		m_1 = \sum_i \eta_i v_i \; , \quad m_2 = \sum_j \Omega_j v_j \;, \quad
		m_3 = \sum_k \Lambda_k v_k \;.
	\end{equation}
	Hence, all fermion masses are independent linear combinations of the different 
	vevs and all Higgs doublets can be responsible for giving mass to all fermions. In practice, models may also lead to Yukawa matrices with ranks less than $3$.
	In this case the singular alignment can still hold and the only new difference
	would be to have some of the constants $\eta_i, \Omega_i, \Lambda_i$ appearing in Eq.~\eqref{eq:SingAlig}
	equal to zero. 
	
	In short, \textit{singular alignment} is
    the very strong Ansatz of choosing Yukawa matrices to be 
    related to the rank $1$ matrices appearing in the SVD.
	Through this alignment, no tree-level FCNC appear for any number of Higgs doublets.
	Let us consider now some explicit examples. 
	
	\subsection{The Two-Fermion Family Case}
\noindent 
	We assume $N$ Higgs doublets for two generations of  charged fermions. In this case,
	the mass matrix is 
	\begin{align}
		{\bf m} = v_1 {\bf y}_1 + \cdots + v_N {\bf y}_N \; .
	\end{align}
	If no symmetry is imposed all Higgs doublets are allowed to couple to our $6$ fermions.
	Therefore, all the Yukawa matrices are rank $2$. Let us implement the singular
	alignment. For this purpose, the SVD of the mass matrix is written as 
	\begin{align}
		{\bf m} = e^{-i\beta_3}\begin{pmatrix}
			c_\alpha e^{-i\beta_1} & -s_\alpha e^{-i\beta_2} \\
			s_\alpha e^{i \beta_2} & c_\alpha e^{i \beta_1}
		\end{pmatrix} \begin{pmatrix}
		m_1 & 0 \\
		0 & m_2
		\end{pmatrix} \begin{pmatrix}
		1 & 0 \\
		0 & 1
		\end{pmatrix} ,
	\end{align}
	where, without any loss of generality, we have chosen to work in the basis where 
	the right-handed fermions have been already transformed and we have explicitly written
	the most general expression for a unitary matrix in two dimensions. The two singular
	matrices are 
	\begin{align}
		{\bf \Delta}_1 = e^{-i\beta_3} \begin{pmatrix}
			c_\alpha e^{-i\beta_1} &0 \\
			s_\alpha e^{i \beta_2} & 0
		\end{pmatrix} \;, \qquad
		{\bf \Delta}_2  = e^{-i\beta_3} \begin{pmatrix}
			0 & -s_\alpha e^{-i\beta_2} \\
			0 & c_\alpha e^{i \beta_1}
		\end{pmatrix} .
	\end{align}
	
	Singularly aligning our Yukawa matrices in flavour space means 
	\begin{align}
		{\bf y}_i = \eta_i {\bf \Delta}_1 + \Omega_i {\bf \Delta}_2 \;,
	\end{align}
	which leads to
	\begin{align}
		{\bf m} = \sum_i \eta_i v_i  {\bf \Delta}_1 + 
		\sum_i \Omega_i v_i  {\bf \Delta}_2 \; .
	\end{align}
	We identify the masses as
	\begin{align} \label{eq:genexp}
		m_1 =  \sum_i \eta_i v_i \qquad \text{ and } \qquad
		m_2 =  \sum_i \Omega_i v_i  \; .
	\end{align}
	
	Regarding FCNC, note that  in the mass basis we have 
	\begin{align}
	 \widetilde{\bf m} = 	{\bf L} {\bf m} = \; m_1 \begin{pmatrix}
		1 & 0 \\
		0 & 0
		\end{pmatrix} +
		m_2 \begin{pmatrix}
		0 & 0 \\
		0 & 1
		\end{pmatrix} .
	\end{align}
	Hence, no FCNC are introduced in this model of  $N>1$ Higgs doublets that couple to all fermions. Also, if matrices of lower rank are obtained through the use of convenient 	symmetries, then our general expressions in Eq.~\eqref{eq:genexp} will still hold
	but with some of the parameters $\eta_i$ or $\Omega_i$  vanishing. 
	
\subsection{\label{sec:AppB}The Three-Fermion Family Case}
\noindent 
The next example deals with three generations and three 
 Higgs doublets. 
To be singularly aligned, each rank $1$ Yukawa matrix
\begin{align}
	{\bf Y}_1  = \eta_1 \begin{pmatrix}
		a_1 & 0 & 0 \\
		a_{2} & 0 & 0 \\
		a_{3} & 0 & 0
	\end{pmatrix} \;, \quad
	{\bf Y}_2 = \Omega_2 \begin{pmatrix}
		0& b_1 & 0 \\
		0& b_{2} & 0 \\
		0& b_{3} & 0
	\end{pmatrix} \;, \quad
	{\bf Y}_3  = \Lambda_3 \begin{pmatrix}
		0 & 0 & c_1 \\
		0 & 0 & c_2 \\
		0 & 0 & c_3
	\end{pmatrix} ,
\end{align}
should be seen as a column vector satisfying unitarity conditions (recall Eq.~\eqref{eq:SingAlig}): 
\begin{align}
\begin{split}
	\langle a| a \rangle =1 \;, \quad \langle b| b \rangle =1\;, \quad \langle c| c \rangle =1 \;, \\
	\langle a| b \rangle =0 \;, \quad \langle a| c \rangle =0\;, \quad \langle b| c \rangle =0 \;. 
\end{split}
\end{align}
Here we have denoted $(a_1, a_2, a_3)^T \equiv | a \rangle$ and similarly
for the other columns.
Notice that the Yukawa couplings should not enter into these expressions.
A practical way to implement all these conditions is to make use of an explicit
parametrization of a unitary matrix. Then, a singularly aligned mass matrix 
could take the form 
\begin{align} \label{eq:SingMassMatrix}
	{\bf M} = e^{i \varphi} \,{\bf T} \begin{pmatrix}
		v_1 \eta_1 c_{\alpha} c_{\gamma} & v_2 \Omega_2 s_{\alpha} c_{\gamma} & v_3 \Lambda_3 s_{\gamma} e^{-i\chi} \\
		-v_1 \eta_1( s_{\alpha} c_{\beta}+c_{\alpha}s_{\beta}s_{\gamma} e^{i\chi}) 
		& v_2 \Omega_2 ( c_{\alpha} c_{\beta}-s_{\alpha}s_{\beta}s_{\gamma} e^{i\chi}) 
		& v_3 \Lambda_3 s_{\beta}c_{\gamma} \\
		v_1 \eta_1 ( s_{\alpha} s_{\beta}-c_{\alpha}c_{\beta}s_{\gamma} e^{i\chi}) 
		& -v_2 \Omega_2 ( c_{\alpha} s_{\beta}+s_{\alpha}c_{\beta}s_{\gamma} e^{i\chi}) 
		& v_3 \Lambda_3 c_{\beta}c_{\gamma}
	\end{pmatrix} {\bf Q} \; .
\end{align}
where ${\bf T}$ and ${\bf Q}$ are diagonal phase matrices with two phases each and
we have used the shorthand notation 
for the sine and cosine functions. Recall that a $3\times 3$ unitary matrix possesses
$6$ complex phases (one of which is global) and 3 real parameters. 
Each column is proportional to a given singular matrix. At last, realize that masses 
and mixing get completely decoupled when singularly aligning the Yukawa matrices. 
Recall in this context that any set of singular vectors corresponding to 
a set of non-degenerate singular values is always orthonormal. 
\\	
	
\subsection{Hierarchical Fermion Masses\label{sec:hie}}
\noindent 
A shared feature among all the charged fermions is that their masses are
hierarchical, 
\begin{align}
    m_1 \ll m_2 \ll m_3 \;.
\end{align}
To theoretically understand this in a $N$HDM with singular alignment, 
see Eq.~\eqref{eq:massesNHDM},
one must understand under what conditions this property gets always realized.
We are not interested in any fine-tuned scenario where through adequate values
for the set of parameters $\{\eta, \Omega, \Lambda \}$ we generate hierarchical
masses, we are assuming that 
$\eta_i, \Omega_i, \Lambda_i = {\cal O}(1)$.
Furthermore, we are actually interested in the minimal number of scalar
doublets necessary to explain all the observed patterns in the fermion masses.
For the moment, notice that one possibility is to couple a single Higgs to each
different flavour with the same electric charge. In this case we have 
	\begin{align}
		m_1 = \eta_1 v_1 \; , \quad m_2 = \Omega_2 v_2 \;, \quad
		m_3 = \Lambda_3 v_3 \;,
	\end{align}
	
It is obvious then that the only way to achieve hierarchical masses with ${\cal O}(1)$ parameters 
is through hierarchical vevs, i.e.\
\begin{align}
    v_3 \gg v_2 \gg v_1 \;.
\end{align}
This fact is connected to the mass-vacuum relation. 

The maximal setup, if neutrinos are assumed as Dirac particles, would require 12 Higgs doublets. 
However, this large number of scalars can be significantly reduced if one notices 
that among the different masses there are majorly 4 (5) mass scales, where 
the (5) corresponds to Dirac neutrino masses. This is what we will deal with in Sec.\ \ref{sec:model}. In case of Majorana neutrinos there are four 
possibilities depending on from which Higgs doublet the Dirac mass matrix of the type-I seesaw mechanism stems. 
We will come back to this point later. Of course, neutrino mass could also be independent of the Higgs doublets. 
	
\subsection{Beyond Singular Alignment}	
\noindent 
If a small amount of flavour violation via neutral mediators is permitted, 
then a less restrictive venue can be obtained through the following conditions: (i) the third Yukawa
matrix for all fermion species is the only rank $1$ matrix and proportional to the third singular matrix,
	\begin{align}
		{\bf Y}_3 = \Lambda_3 {\bf \Delta}_{3} \;;
	\end{align}
(ii) the first and second Yukawa matrices are no longer proportional to the singular matrices, 
so they may in general produce FCNC; 
(iii) however, to produce a hierarchy between the first and second generation, the second Yukawa matrix should be at most
	rank $2$ and have no contributions to the first family masses;  
(iv) the first Yukawa matrix can be rank $3$, $2$ or $1$. 
In other words, the three Yukawa matrices should imply the sequential symmetry breaking chain 
\begin{align}
	U(3)^3 \xrightarrow[{\bf Y}_{f,3}]{} U(2)^3 \xrightarrow[{\bf Y}_{f,2}]{}  U(1)^3 \xrightarrow[{\bf Y}_{f,1}]{} U(1)_F \;,
\end{align}
where $F$ might either be baryon or lepton number. The introduction of flavour violation as allowed by the
two lightest families means no risk as this set of flavour transitions will be sequentially suppressed
by the approximately conserved symmetries at each step.

\subsection{Radiative Stability} \label{ssection:RadStab}
\noindent
In the absence of a specific symmetry protection, 
one-loop quantum corrections may induce misalignment in the different singularly aligned
Yukawa matrices and bring about FCNC's at the loop level. 
It is important to know if this effect is small and 
compatible with current experimental constraints. 
The study of this issue can be directly related to the work of Ref.~\cite{Penuelas:2017ikk}
wherein the issue of radiative stability was investigated for 
the most generalized Yukawa aligned-like form given by
\begin{align} \label{eq:GYA}
	{\bf Y}_i = \frac{{\bf \Xi}_i}{v_i} (m_1 {\bf \Delta}_1 + m_2 {\bf \Delta}_2 + m_3 {\bf \Delta}_3) \;,
\end{align}
where ${\bf \Xi}_i$ is a complex $3\times 3$ matrix subject to the condition
\begin{align} \label{eq:GeneralYukAlig}
	{\bf L} {\bf \Xi}_i {\bf L}^\dagger = \text{diag}(\zeta^{(1)}_i, \zeta^{(2)}_i, \zeta^{(3)}_i) \;.
\end{align}
This generalized Yukawa-alignment means breaking flavour universality. Notice that the normal 
Yukawa-alignment, Eq.~\eqref{eq:NormalYukAlig}, can be recovered when all diagonal elements in the
r.h.s of Eq.~\eqref{eq:GeneralYukAlig} are equal (flavour universal). 
In Ref.~\cite{Penuelas:2017ikk}, it was shown that the induced misalignment is a quite 
small effect, as the initial alignment in the multi-Higgs Lagrangian has some residual flavour symmetries,
which tightly limit the type of FCNC operators that can be generated at higher orders.
This can be easily understood as the Yukawa alignment 
is a linear realization of the minimal flavour violation hypothesis~\cite{Kagan:2009bn} and could be derived
from it~\cite{Buras:2010mh}. This hypothesis states that
the only source of flavour breaking should come from the Yukawa matrices, even in the
presence of new particles and interactions~\cite{DAmbrosio:2002vsn,Hall:1990ac,Chivukula:1987py,Buras:2000dm}.   

The previous discussion also applies to the Singular Alignment as it is possible to show that it is equivalent to the generalized Yukawa-alignment
via substitution in Eq.~\eqref{eq:GYA} of the relations
\begin{align}
	\zeta^{(1)}_i = \frac{\eta_i v_i}{m_1} \;, \quad
	\zeta^{(2)}_i = \frac{\Omega_i v_i}{m_2} \;, \quad
	\zeta^{(3)}_i = \frac{\Lambda_i v_i}{m_3} \;.
\end{align}
Therefore, the ansatz of singularly aligning Yukawa
matrices in flavour space, in order to avoid FCNC's at tree level, has a sufficiently small
misalignment, induced by one-loop quantum corrections,  consistent with all
known phenomenological tests. 

\section{\label{sec:model}The Minimal Setup: a 4HDM}
\begin{figure}[t]
\begin{center}
		\includegraphics[scale=0.34]{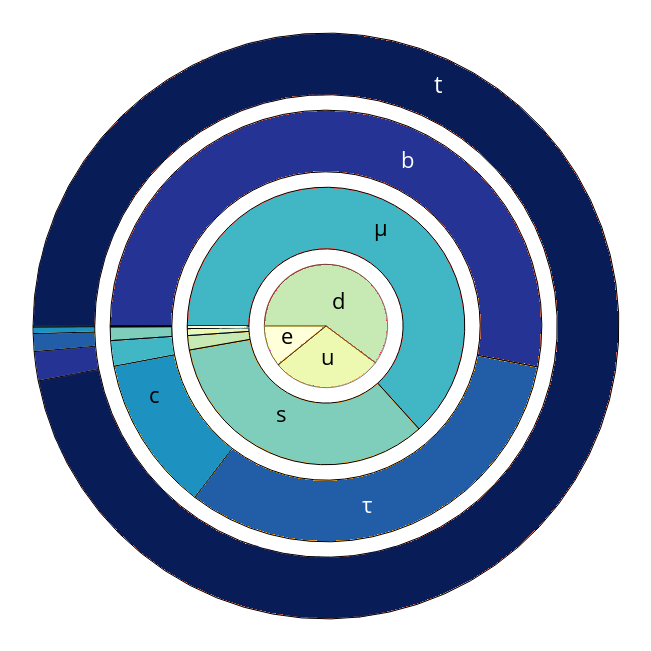}
	\caption{This set of sector charts illustrates the hierarchy among the set of charged
	fermion masses, $\{m_t,m_b,m_\tau,m_c,m_\mu,m_s,m_d,m_u,m_e\}$. 
	Each mass has its own color, the darker the color the heavier the mass.
	The outer ring  shows how $m_t \gg \{m_f\} \; (f\neq t)$.
	The following one has the same set of masses without its largest previous
	contribution, i.e.\ the top quark mass. Here the fermions in the set $\{ m_b,m_\tau,m_c \}$ are 
	of similar mass while the other fermions are much lighter. The same logic continues with
	the other two smaller rings.
	The major contributions to each ring imply four different groups: 
	$\{ m_t\}$, $\{ m_b,m_\tau,m_c \}$,
	$\{ m_\mu,m_s \}$, and $\{ m_d,m_u,m_e \}$, illustrating a similar scale, 
	$174 \text{ GeV}$, $1 \text{ GeV}$, $0.1 \text{ GeV}$, and $0.001 \text{ GeV}$, 
	for each group, respectively.  If neutrinos were considered as Dirac
	particles their masses would require their
	own scale (ring), $m_\nu \lesssim  \text{eV}$. 
	}
	\label{fig:4vevs}
\end{center}
\end{figure}

\noindent
Now we discuss a 4HDM which takes into account that among the measured fermion masses four 
different sets can be identified: 
$\{ m_t\}$, $\{ m_b,m_\tau,m_c \}$, $\{ m_\mu,m_s \}$ and $\{ m_d,m_u,m_e \}$. 
Within each set  the masses are within one order of magnitude. This fact is depicted in Figure~\ref{fig:4vevs}. 
We will introduce four Higgs doublets $\Phi_t$, $\Phi_b$, $\Phi_\mu$ and $\Phi_d$, which are responsible for the masses in their respective sets\footnote{As mentioned before, in principle neutrinos provide a different set of masses $\{ m_{\nu 1},m_{\nu 2},m_{\nu 3} \}$.}. 
The corresponding mass-vacuum-like relation in analogy to Eq.\ (\ref{eq:mvr}) would take the form
\begin{align} \label{eq:4hdm-massvacuum}
	v_t^2 + v_b^2 + v_\mu^2 +v_d^2 = v_\text{EW}^2 \;.
\end{align}

The model can be constructed by first imposing fields to transform under the symmetry 
$Z_2 \times Z'_2 \times Z''_2$, as shown in Table~\ref{tab:4HDM}. 
\begin{table}
\centering
{\bf SCALAR SECTOR}\\
\begin{tabular}{c|cccc}
  \toprule[0.1em] 
  & \(\Phi_t\) & \(\Phi_b\) &  \(\Phi_\mu\)  & \(\Phi_d\)   \\
  \hline
  $Z_2$  & \(+\) & $-$ & $-$ & $-$  \\
  $Z'_2$ & \( + \) & $+$ & $-$ & $-$ \\
  $Z''_2$ & \( + \) & $+$ & $+$ & $-$ \\
  \bottomrule[0.1em]
\end{tabular}\\ \vspace{.51 cm}
{\bf QUARK SECTOR}\\
\begin{tabular}{c|ccc|ccc|ccc}
  \toprule[0.1em]
  & \(Q_{3L}\) & \(Q_{2L}\) & \(Q_{1L}\) 
  & \(u_{3R}\) & \(u_{2R}\) & \(u_{1R}\) 
  & \(d_{3R}\) & \(d_{2R}\) & \(d_{1R}\) \\
  \hline
  $Z_2$  & \(+\) & $+$ & $+$ & $+$ & $-$ & $-$ & $-$ & $-$ & $-$ \\
  $Z'_2$ & \( +\) & $+$ & $+$ & $+$ & $+$ & $-$ & $+$ & $-$ & $-$ \\
  $Z''_2$ & \( +\) & $+$ & $+$ & $+$ & $+$ & $-$ & $+$ & $+$ & $-$ \\
  \bottomrule[0.1em]
\end{tabular} \\ \vspace{.51 cm}
{\bf LEPTON SECTOR}\\
\begin{tabular}{c|ccc|ccc|ccc}
  \toprule[0.1em]
  & \(E_{3L}\) & \(E_{2L}\) & \(E_{1L}\) 
  & \(n_{3R}\) & \(n_{2R}\) & \(n_{1R}\) 
  & \(e_{3R}\) & \(e_{2R}\) & \(e_{1R}\) \\
  \hline
  $Z_2$  & \(+\) & $+$ & $+$ & $-$ & $-$ & $-$ & $-$ & $-$ & $-$ \\
  $Z'_2$ & \( +\) & $+$ & $+$ & $-$ & $-$ & $-$ & $+$ & $-$& $-$ \\
  $Z''_2$ & \( +\) & $+$ & $+$ & $-$ & $-$ & $-$ & $+$ & $+$& $-$ \\
  \bottomrule[0.1em]
\end{tabular}
\caption{Charge assignment under the discrete flavour symmetry group $Z_2 \times Z'_2 \times Z''_2$ 
for the different scalar and fermion fields in the 4HDM with rank one Yukawa matrices. 
This extension to the SM comprises $3$ right-handed neutrinos and $3$ new Higgs fields.
\label{tab:4HDM} }
\end{table}
The Yukawa Lagrangian implied by the charge assignment is 
\begin{align}
\begin{split}
	- {\cal L}_Y = \sum_{i=1}^3 
	\left[ y_{i}^t \overline{Q}_{L,i} \widetilde{\Phi}_t u_{3R}  
	+ y_{i}^c \overline{Q}_{L,i} \widetilde{\Phi}_b u_{2R} 
	+ y_{i}^u \overline{Q}_{L,i} \widetilde{\Phi}_d u_{1R} 
	+ y_{i}^b \overline{Q}_{L,i} {\Phi}_b d_{3R}  \right.\\\left.
    + y_{i}^\mu \overline{Q}_{L,i} {\Phi}_\mu d_{2R} 
	+ y_{i}^d \overline{Q}_{L,i} {\Phi}_d d_{1R} 
	+ y_{i}^\tau \overline{E}_{L,i} {\Phi}_b e_{3R}  
	+ y_{i}^\mu \overline{E}_{L,i} {\Phi}_\mu e_{2R} \right.\\\left.
    + y_{i}^e \overline{E}_{L,i} {\Phi}_d e_{1R}
	+ \overline{E}_{L,i} {\bf Y}^\nu_D \widetilde{\Phi}_d n_{3R} 
	+ \frac 12 {\cal M}^\nu_{ij} \overline{n_{R,i}^c} n_{R,j}
	+ \text{ H.c.}
	\right]
\end{split}
\end{align}
The way in which we have employed the charge assignment to couple fermions with Higgs doublets
has given us a model where all Yukawa matrices for the charged fermions are rank $1$. 
For example, in the up-quark sector we have ${\bf M}_{\rm up} = v_d {\bf Y}_u + v_b {\bf Y}_c + v_t {\bf Y}_t$, with 
\begin{align}
{\bf Y}_{u}  = \begin{pmatrix}
     y^{u}_1 & 0 & 0\\
     y^{u}_2 & 0 & 0\\
     y^{u}_3 & 0 & 0
    \end{pmatrix} \;, \quad
 {\bf Y}_{c} = \begin{pmatrix}
    0 &  y^{c}_1 & 0 \\
    0 &  y^{c}_2 & 0\\
    0 &  y^{c}_3 & 0
    \end{pmatrix} \;, \quad
    {\bf Y}_{t}  = \begin{pmatrix}
    0 & 0 & y^{t}_1 \\
    0 & 0 & y^{t}_2 \\
    0 & 0 & y^{t}_3
    \end{pmatrix} .
\end{align}
Similar expressions can be given for the other fermion species. 
Notice we are employing a conventional notation for the Yukawa couplings, $y^f_i$,
in order to distinguish at this point generic Yukawa matrices from
those which have been singularly aligned.

Now, to singularly align these matrices, we demand that each column should be 
given by a single singular matrix (in order to have a hierarchy of masses with order one Yukawa couplings, cf.\ Section \ref{sec:hie}), in our up-type example this means: 
\begin{align}
\begin{split}
    {\bf Y}_{u}  = \eta_{u} {\bf \Delta}_{u} \;, \qquad 
    {\bf Y}_{c} = \Omega_{c} {\bf \Delta}_{c} \;, \qquad
    {\bf Y}_{t}  = \Lambda_{t} {\bf \Delta}_{t} \; .
\end{split}
\end{align}
The explicit form of these singular matrices was given in Section~\ref{sec:AppB}, they 
correspond to one of the three columns in Eq.~\eqref{eq:SingMassMatrix}. 
We can also write them as $\Delta_{u,c,t} = {\bf L}^\dagger P_{1,2,3} \,{\bf R}$, see  the discussion around Eq.\ (\ref{eq:11}).

The model presented here arranges that a certain Higgs doublet will couple to a given set of fermions, even if they  possess different electric charge. All corresponding Yukawa matrices will already be rank $1$. 
Through the special requirement that Yukawa matrices should be \textit{singularly} 
aligned in flavour space, as discussed in Sec.\ \ref{sec:SA}, it is possible to avoid 
flavour violation at tree-level. 

The model allows to reproduce fermion mixing, as shown in Appendix~\ref{sec:appMixing}. 
Neutrino masses are generated via the type-I seesaw mechanism. We have associated 
the three right-handed neutrinos to the Higgs doublet $\Phi_d$. This implies that via the 
type-I seesaw mechanism the heavy neutrino mass scale 
${\cal M}$ should be around PeV, where we have assumed that $m_D \simeq \langle \Phi_d^0 \rangle  \simeq 10$ MeV and $m_\nu \simeq m_D^2 /{\cal M} \simeq 0.1$ eV.  
The contributions to some lepton-flavor-violation processes coming from the admixture
of the heavy right-handed neutrinos with the left-handed ones can already be estimated 
via the standard formulae of type-I seesaw models~\cite{Ilakovac:1994kj}. 
This calculation is greatly simplified in the limit ${\cal M} \gg M_W$, which is our case\footnote{A complete analysis of the model in the lepton sector is outside the scope of this paper and will be
presented elsewhere.}.
The following upper bound to various processes of interest may be obtained: ${\cal B}^\text{th}_r(\mu \rightarrow e \gamma) < 10^{-14}$,
${\cal B}^\text{th}_r(\mu \rightarrow 3e) < 10^{-18}$, ${\cal B}^\text{th}_r(\tau \rightarrow 3\mu ) < 10^{-7}$,
and ${\cal B}^\text{th}_r(\tau \rightarrow \mu \gamma) < 10^{-4}$.
Notice how, in general, these numbers will still get suppressions by small mixing-like angles of the order of $m_D / {\cal M} \sim 10^{-8}$ times order one numbers (at most) arising from corresponding form factors. The present experimental upper limits on these decays at $90$\% C.L. are given by: 
	${\cal B}^\text{exp}_r(\mu \rightarrow e \gamma) < 4.2 \times 10^{-13}$~\cite{TheMEG:2016wtm},
	${\cal B}^\text{exp}_r(\mu \rightarrow 3e) < 10^{-12}$~\cite{Bellgardt:1987du},
	${\cal B}^\text{exp}_r(\tau \rightarrow 3\mu ) < 4.6 \times 10^{-8}$~\cite{Aaij:2014azz} and
	${\cal B}^\text{exp}_r(\tau \rightarrow \mu \gamma) < 3.3 \times 10^{-8}$~\cite{Aubert:2009ag}.
The smallness of the estimated branching ratios is of no surprise, as the high-scale type-I seesaw is known for giving 
very suppressed rates, see for example~\cite{Hambye:2013jsa} and references therein.
On the other hand, possible contributions coming from the scalar mediators
at the loop level can also be expected to be sufficiently small and consistent with 
phenomenological tests as suggested by analyses of the minimal lepton-flavour violation hypothesis~\cite{Cirigliano:2005ck,Cirigliano:2006su} and as discussed in Section~\ref{ssection:RadStab}. Notice, however, that this set of flavour-violating processes
have a strong dependence in the ratio between the two scales 
$(\Lambda_\text{LN} / \Lambda_\text{LFV})^4$, where the first and second one correspond to
the scale where lepton number (LN) is broken and lepton-flavour-violation (LFV) is produced. 
Therefore, if a large hierarchy exists between these two scales one may obtain observable
effects. In our case, we may estimate this ratio as 
$(v_d {\cal M}/v_{\tau}^2)^2 \sim 10^{8}$ which is still sufficiently small. For example,  
after substitution in ${\cal B}_r^\text{th} (\mu \rightarrow e \gamma) = 1.6 \times 10^{-24}\,(\Lambda_\text{LN} / \Lambda_\text{LFV})^4$ we obtain ${\cal B}_r^\text{th} (\mu \rightarrow e \gamma)\sim 10^{-16}$, where the previous relation was taken from Ref.~\cite{Cirigliano:2005ck}. Hence, we see that for the particular purposes of this work the rates for LFV
processes are expected to be in agreement with the current upper bounds.

\subsection{The Scalar Potential}
\noindent 
The most general, renormalizable and gauge invariant scalar potential of the model is 
$V = V_0 + V_\text{soft}$, where 
\begin{align}
\begin{split}
    V_0 & = \sum_ {a} \left[ \mu_a^2 (\Phi^\dagger_a \Phi_a) + \frac{\lambda_a}{2} (\Phi^\dagger_a \Phi_a)^2 \right] 
    + \sum_{a \neq b} X_{ab} (\Phi^\dagger_a \Phi_a)  (\Phi^\dagger_b \Phi_b) \\
	&+ \sum_{a \neq b} Y_{ab} (\Phi^\dagger_a \Phi_b)  (\Phi^\dagger_b \Phi_a)
	+ \sum_{a \neq b} \frac{Z_{ab}}{2} \left[ (\Phi^\dagger_a \Phi_b)^2 + (\Phi^\dagger_b \Phi_a)^2 \right] .
\end{split}
\end{align}
Here ${a,b=t,b,\mu,d}$, and for the sake of simplicity 
we are assuming all couplings to be real. The term 
$V_0$ is invariant under $Z_2 \times Z'_2 \times Z''_2$, whereas 
$V_\text{soft}$ includes different soft-breaking terms ($V_\text{soft} \ll V_0$), see below. 

In order to generate a hierarchy among the vevs we choose the particular case where 
\begin{align}
    \mu_t^2 <0 \quad \quad \text{and} \quad \quad \mu_{b,\mu,d}^2 > 0 \;,
\end{align}
such that the only Higgs acquiring a vev is $\Phi_t$: 
\begin{align} \label{eq:vt}
    \frac{\partial V_0}{\partial \Phi_t} \Bigg{|}_\text{min} = 0 \quad \Rightarrow \quad 
    v_t = \sqrt{\frac{-\mu_t^2}{\lambda_t}} \;.
\end{align}
We are following the convention  
$\langle \Phi_t^0 \rangle = v_t$ (which differs from $\langle \Phi_t^0 \rangle = v_t/\sqrt{2}$).
As $\Phi_t$ has no charge under any of the three Abelian symmetries, see Table 
\ref{tab:4HDM}, its vev preserves the symmetry. 
Equivalently, the symmetries are protecting the other scalars from acquiring a vev.
Thereafter, through the following subset of soft-breaking terms, 
\begin{align}
\label{eq:VSoft}
    V_\text{soft}  \supset \mu_{tb}^2 \left( \Phi_t^\dagger \Phi_b + \Phi_b^\dagger \Phi_t  \right) +
     \mu_{b\mu}^2 \left( \Phi_b^\dagger \Phi_\mu + \Phi_\mu^\dagger \Phi_b  \right) 
      +  \mu_{\mu d}^2 \left( \Phi_\mu^\dagger \Phi_d + \Phi_d^\dagger \Phi_\mu  \right),
\end{align}
where $\mu_{ab}^2 \ll v_t^2,\mu^2_a$,  we induce  vevs for the 
other three Higgs doublets. To be more specific, 
the particular choice of soft-breaking terms is motivated by  the fact
that each of them will only break a particular piece of the whole symmetry. That is, 
$(\Phi^\dagger_t \Phi_b + \Phi^\dagger_b \Phi_t)$,
$(\Phi^\dagger_b \Phi_\mu + \Phi^\dagger_\mu \Phi_b)$, and
$(\Phi^\dagger_d \Phi_\mu + \Phi^\dagger_\mu \Phi_d)$ only break 
$Z_2$, $Z'_2$, and $Z''_2$,
correspondingly.
Therefore, once the EW symmetry is spontaneously broken, the first soft-breaking term
will induce a vev to $\Phi_b$ which in return will induce a vev to $\Phi_\mu$ until
finally reaching $\Phi_d$. 
It is possible to show that within this limit the 
minimization conditions are satisfied if the vevs are given as 
\begin{align}\label{42}
    v_b \simeq \frac{- v_t \mu^2_{tb}}{(XYZ)_{tb} v_t^2 + \mu_b^2} \;, \qquad
    v_\mu \simeq \frac{- v_b \mu^2_{b\mu}}{ (XYZ)_{t \mu} v_t^2 + \mu_\mu^2}\;, \qquad
    v_d \simeq \frac{- v_\mu \mu^2_{\mu d}}{(XYZ)_{t d} v_t^2 + \mu_d^2}\;,
\end{align}
together with  Eq.~\eqref{eq:vt} and where $(XYZ)_{ab} = X_{ab} + Y_{ab} + Z_{ab}$
and $\mu^2_{ab} < 0$.
By virtue of this choice, the vevs are naturally small and obey the desired hierarchy 
\begin{align}
v_t^2 \gg v_b^2 \gg v_\mu^2 \gg v_d^2 \;.    
\end{align}
For example, with $v_t \simeq 174 \text{ GeV}$, 
$\mu_{b,\mu,d} \sim 200 \text{ GeV}$,
$|\mu_{(tb),(b\mu), (\mu d)}| \sim 35 \text{ GeV}$ one finds $v_b \sim 1 \text{ GeV}$, 
$v_\mu\sim 0.1 \text{ GeV}$ and $v_d \sim 0.001 \text{ GeV}$. 

\begin{figure*}[ht]
\begin{center}
		\includegraphics[scale=0.33]{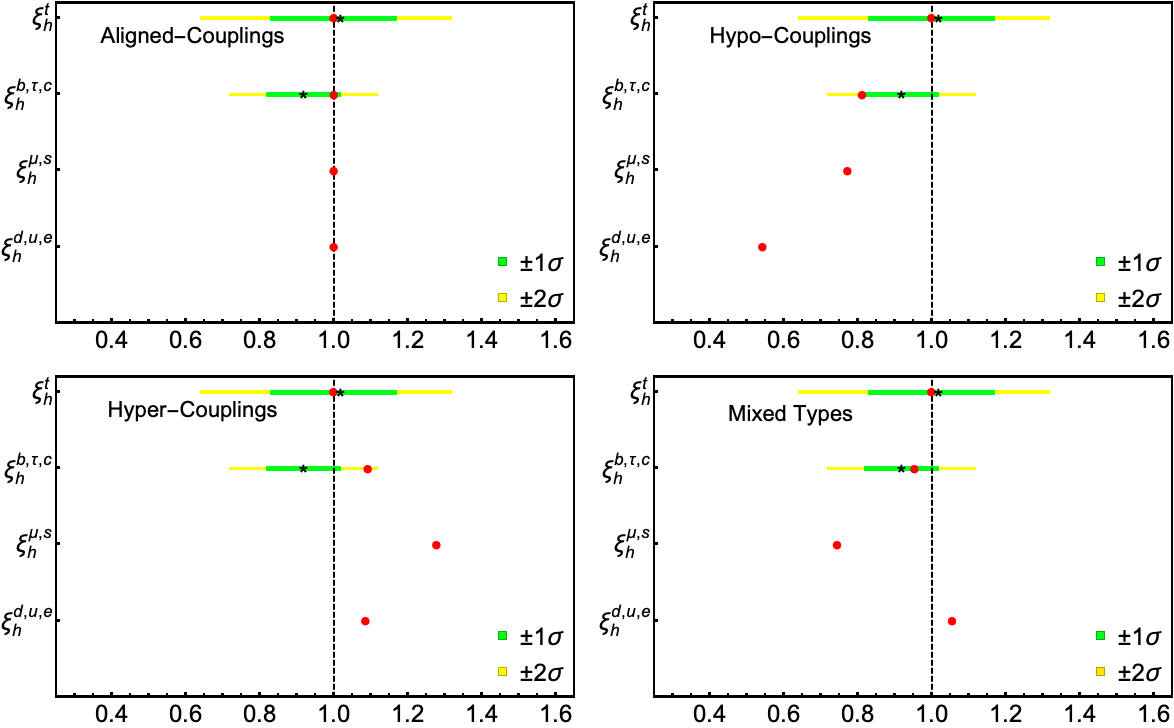}
	\caption{The red dots are predictions for the modified couplings of fermions in the sets 
	$\{ m_t \}$, $\{ m_b,m_\tau,m_c \}$, $\{ m_\mu,m_s \}$, 
and $\{ m_d,m_u,m_e \}$ to the SM-like scalar. The four different benchmark scenarios are defined in Table \ref{tab:NumExam}. The vertical dashed line 
	corresponds to the SM expectation. The black star is the central value of the measurement with green (yellow) bands being the $1\sigma\, (2\sigma)$ ranges. For simplicity the measured couplings of the Higgs to the bottom quark and tau
	lepton have been merged here into a single one, $\kappa_{b,\tau} = 0.92 \pm 0.10$.}
	\label{fig:couplings}
\end{center}
\end{figure*}

\subsection{Fermionic Couplings to the SM-like Higgs}
\noindent 
The introduction of the soft breaking terms (Eq.~(\ref{eq:VSoft})) in the Higgs potential 
will produce a small mixing among the four Higgs doublets. For the moment, let us focus on the neutral scalars. We assume all parameters in the scalar potential to be real. Through this
choice we consider it to be $CP$-symmetric.  Hence, no admixture between the
real and imaginary components of the neutral fields is allowed as 
they have definite $CP$ quantum numbers. To compute their couplings to all 
fermions we start from the Yukawa Lagrangian in the mass basis which is written as 
\begin{align} \label{eq:MassBasisFerHigg}
\begin{split}
    - {\cal L}_Y & = y_t \bar{t} t \left( v_t + \frac{\phi_t}{\sqrt{2}} \right) 
      + \sum_{f=b,\tau,c} y_f \bar{f} f \left( v_b + \frac{\phi_b}{\sqrt{2}} \right) \\
    & + \sum_{f=\mu,s} y_f \bar{f} f \left( v_\mu + \frac{\phi_\mu}{\sqrt{2}} \right) 
     + \sum_{f=d,u,e} y_f \bar{f} f \left( v_d + \frac{\phi_d}{\sqrt{2}} \right),
\end{split}
\end{align}
where we have changed our notation $\{\eta,\Omega,\Lambda\}$ to the conventional one, $y_f$. We can bring the $CP$-even scalar sector to its mass basis via 
\begin{align}
	\begin{pmatrix}
		\phi_t \\
		\phi_b \\
		\phi_\mu \\
		\phi_d
	\end{pmatrix} = 
	{\cal R}^T\begin{pmatrix}
		h_0 \\
		h_1\\
		h_2 \\
		h_3
	\end{pmatrix}  ,
\end{align}
where ${\cal R}$ is an orthogonal matrix, ${\cal R}^T {\cal R} = {\cal R}{\cal R}^T = 1_{4 \times 4}$,
and $h_0$ is the lightest state with a mass of $m_{h_0} \simeq 125 \text{ GeV}$.
Now, in order to find out how fermions  couple to the SM-like Higgs, $h_0$, we substitute $\phi_k = {\cal R}_{1k}h_{0} $ in Eq.~\eqref{eq:MassBasisFerHigg} to obtain 
\begin{align}
    -{\cal L}_Y \supset \sum_{f} \frac{m_f}{(246 \text{ GeV})} \xi^f_h \bar{f}f h_0 \;.
\end{align}
We can define the following four classes of fermion-scalar couplings: 
\begin{align}
\begin{split}
    \xi^{t}_h & = \frac{{\cal R}_{11}}{\cos \alpha_1} \;, \\
    \xi^{b,\tau,c}_h & = \frac{{\cal R}_{12}}{\sin \alpha_1 \cos \alpha_2} \;, \\
    \xi^{\mu,s}_h & = \frac{{\cal R}_{13}}{\sin \alpha_1 \sin \alpha_2 \cos \alpha_3} \;, \\
    \xi^{d,u,e}_h & = \frac{\sqrt{1-\sum_j {\cal R}_{1j}^2}}{\sin \alpha_1 \sin \alpha_2 \sin \alpha_3} \;.
\end{split}
\end{align}
The angles $\alpha_i$ in these relations are 
\begin{align}
    \sin \alpha_1 & = \sqrt{\frac{v_b^2+v_\mu^2+v_d^2}{v_t^2+v_b^2+v_\mu^2+v_d^2}} \;, \\
    \sin \alpha_2 & = \sqrt{\frac{v_\mu^2+v_d^2}{v_b^2+v_\mu^2+v_d^2}}  \;,\\
    \sin \alpha_3 & = \sqrt{\frac{v_d^2}{v_\mu^2+v_d^2}} \;.
\end{align}
We note an attractive and testable feature of the model, namely that the 
couplings between fermions and the SM-like Higgs are modified in the same way for each set. That is, the couplings of the sets $\{ m_t \}$, 
$\{ m_b,m_\tau,m_c \}$, $\{ m_\mu,m_s \}$, 
and $\{ m_d,m_u,m_e \}$ are changed with respect to the SM-case by the same amount for each set, see  Fig.~\ref{fig:couplings}. The coupling to the top quark is always essentially SM-like, $\xi^{t}_h = 1$. This is understood because ${\cal R}_{11}$ and $\cos \alpha_1$ are both very close to $1$, which is caused by the vev hierarchy $v_t \gg v_{b,\mu,s}$.

Notice that, even though the mixing ${{\cal R}_{ik}}$ in all cases is proportional to the soft-breaking parameters,
the implied smallness in $|{{\cal R}_{ik}}|$ may be compensated by $\alpha_i \ll 1$, and therefore, in general, $\xi_{h}^f$ should not be expected to be small.  
In fact, within this scenario we can have four different  possibilities: (i) hyper-couplings with $\xi^f_h > 1$, aligned couplings with $\xi^f_h = 1$, 
hypo-couplings with $\xi^f_h < 1$, and a mixture of any of these (the $\xi^f_h$ can even be negative). 
One has to confront the couplings in this model 
with present measurements of Higgs couplings. 
We adopt the following numbers from combined fits of data taken at $\sqrt{s} = 13 \text{ TeV}$~\cite{Sirunyan:2018koj}: 
\begin{align}
    \kappa_Z & = -0.87^{+0.08}_{-0.08} \;, \quad
    \kappa_W = -1.00^{+0.09}_{-0.00} \;, \quad
    \kappa_t  = 1.02^{+0.19}_{-0.15} \;, \\
    \kappa_\tau & = 0.93^{+0.13}_{-0.13} \;, \quad
    \kappa_b  = 0.91^{+0.17}_{-0.16} \;, \quad
    \kappa_\mu = 0.72^{+0.50}_{-0.72} \;.
\end{align}
No useful information about the couplings to first and second generation fermions exist, except for the muon, where the uncertainties are nevertheless very large.  
In our case $\kappa_{Z,W}$ can be reproduced as in any multi-Higgs doublet model. The values of 
$\kappa_{t,\tau,b}$ need to be compared with our $\xi_{h}^f$, which is what the plots in Fig.\ \ref{fig:couplings} do for the four benchmark scenarios to be discussed next\footnote{Since the experimentally allowed range for the muon coupling is quite large we do not include it in the plots.}. 

\begin{table*}[ht]
\centering
\begin{tabular}{c|cccc}
\toprule[0.1em]\toprule[0.1em]
& Aligned & Hypo & Hyper & Mixed \\
\midrule[0.1em]
$\mu_t$ & 88.5 & 88.6 & 88.6& 88.5 \\
$\mu_b$ & 394.8 & 327.3 & 354.0 & 381.3\\
$\mu_\mu$ & 421.3 & 389.1 & 375.0 & 439.5 \\
$\mu_d$ & 496.1 & 426.4 & 378.5 & 468.2 \\ \hline
$|\mu_{tb}|$ & 43.7& 37.3& 38.2& 40.6 \\
$|\mu_{b\mu}|$ & 57.2& 53.9& 51.0& 69.5\\
$|\mu_{\mu d}|$ & 66.1& 61.7& 51.8& 68.3\\ \hline
$\lambda_t$ & 0.26& 0.26& 0.26& 0.26\\
$\lambda_b$ & 1.53 & 0.18& 0.76& 0.34 \\
$\lambda_\mu$ & 0.96& 1.85& 1.41& 0.99\\
$\lambda_d$ & 0.57& 1.49& 0.25& 1.00\\ \hline
$X_{tb}$ & $1.58$ & $0.28$ & $1.43$ & $1.19$ \\
$X_{t\mu}$ & $-0.14$ & $1.59$ & $0.62$ & $0.92$ \\
$X_{td}$ & $1.61$ & $-0.44$ & $1.16$ & $1.13$ \\
$X_{b\mu}$ & $1.04$ & $1.30$ & $1.20$ & $0.77$ \\
$X_{bd}$ & $-0.24$ & $1.45$ & $0.29$ & $0.14$ \\
$X_{\mu d}$ & $1.39$ & $-1.05$ & $-0.50$ & $0.01$ \\ \hline
$Y_{tb}$ & $-0.31$ & $1.94$ & $-0.79$ & $1.44$ \\
$Y_{t\mu}$ & $-0.14$ & $0.78$ & $-0.77$ & $-0.02$ \\
$Y_{td}$ & $0.74$ & $1.16$ & $1.25$ & $0.70$ \\
$Y_{b\mu}$ & $0.04$ & $-1.92$ & $-0.09$ & $-0.30$ \\
$Y_{bd}$ & $1.78$ & $0.40$ & $0.41$ & $-1.85$ \\
$Y_{\mu d}$ & $1.47$ & $-0.07$ & $-1.69$ & $1.63$ \\ \hline
$Z_{tb}$ & $-1.01$ & $-1.61$ & $-0.55$ & $-2.27$ \\
$Z_{t\mu}$ & $0.52$ & $-2.07$ & $0.06$ & $0.05$ \\
$Z_{td}$ & $-2.00$ & $0.23$ & $-1.59$ & $-2.57$ \\
$Z_{b\mu}$ & $-0.87$ & $0.39$ & $-0.17$ & $-0.90$ \\
$Z_{bd}$ & $-2.45$ & $0.01$ & $-1.03$ & $0.57$ \\
$Z_{\mu d}$ & $-2.09$ & $0.25$ & $2.39$ & $-2.72$ \\
\bottomrule[0.1em]\bottomrule[0.1em]
\end{tabular}
\caption{Four sets of numerical values giving rise to four different benchmark scenarios as discussed
in the text. The $\mu_a$ and $\mu_{ab}$ parameters are in GeV while the rest have no units. 
All parameters have been assumed to be real.
\label{tab:NumExam} }
\end{table*}

\begin{table*}[ht]
\centering
\begin{tabular}{c|cccc}
\toprule[0.1em] \toprule[0.1em]
& Aligned & Hypo & Hyper & Mixed \\
\midrule[0.1em]
$\xi_h^t$       & 1.00   & 1.00 & 1.00 & 1.00\\
$\xi_h^{b,\tau,c}$ & 1.00& 0.81 & 1.09 & 0.95 \\
$\xi_h^{\mu,s}$ & 1.00   & 0.77 & 1.28 & 0.74\\
$\xi_h^{d,u,e}$ & 1.00   & 0.54 & 1.08 & 1.05\\
\hline
$m_{h0}^0$ & 125& 125 & 125 & 125\\
$m_{h1}^0$ & 404& 354 & 357 & 395\\
$m_{h2}^0$ & 430& 401 & 372 & 443\\
$m_{h3}^0$ & 507& 459 & 410 & 473\\
\hline
$m_{A1}^0$ & 391 & 443&  366 & 467\\
$m_{A2}^0$ & 474 & 472&  402 & 542\\
$m_{A3}^0$ & 615 & 535&  514 & 594\\
\hline
$M^\pm_1$ & 415& 340& 398& 425\\
$M^\pm_2$ & 452& 410& 411& 470\\
$M^\pm_3$ & 543& 447& 423& 504\\ 
\bottomrule[0.1em] \bottomrule[0.1em]
\end{tabular}
\caption{Outputs for each of the four numerical benchmark scenarios. Scalar masses are given in $\text{GeV}$.
\label{tab:NumExam2} }
\end{table*}

\subsection{Numerical Examples}
\noindent 
A thorough analysis of the Higgs potential is beyond the scope of this work, nevertheless, we will present four numerical benchmark scenarios.  They obey the following conditions and constraints: 
\begin{itemize}
    \item Bounded from below conditions: 
    \begin{align}
        \lambda_{t,b,\mu,d} \geq 0 \;, \qquad 
        X_{ab} \geq - \sqrt{\lambda_a \lambda_b} \;,
    \end{align}
    where $ab = tb,t\mu,td,b\mu,bd,\mu d$. These two conditions are
    necessary but not sufficient. 
    
    \item Unitarity (and perturbativity) bounds:
    \begin{align} \label{eq:unitarity}
0 < \lambda_a \lesssim 2 \;, \quad -4 \lesssim (XYZ)_{ab} \lesssim 2 \;,\quad
|X_{ab}| \lesssim 3 \;, \quad |Y_{ab}| \lesssim 3 \;, \quad |Z_{ab}| \lesssim 3\;, 
    \end{align}
    where again  $(XYZ)_{ab} = X_{ab} + Y_{ab} + Z_{ab}$. 
    We have numerically extracted these relations via the K-matrix
    formalism~\cite{Wigner:1946zz,Wigner:1947zz} as done in~\cite{Bell:2016obu,Bell:2016ekl}.
    
    \item Vacuum stability: 
    \begin{align}
    X_{ta}  > -\frac{\mu^2_a}{v_t^2} \;, \qquad 
    (X+Y-Z)_{ta}  > -\frac{\mu^2_a}{v_t^2} \;, 
    \end{align}
    where $(a=b,\mu,d)$.
    This set of conditions was computed from
    the requirement that the squared mass matrices for the charged scalars and pseudo-scalars 
    should be positive definite, for further details see  Appendix~\ref{sec:appConditions}.
    
    \item Contributions to the $\rho$ parameter:
    \begin{align}
        \Delta \rho = 0.0005 \pm 0.0005 \; (\pm \, 0.0009) \;,
    \end{align}
    that is, it should be consistent with 
    the maximum allowed deviation from the SM-expectation~\cite{Tanabashi:2018oca}. The 
    first and second uncertainty originates whether the oblique parameter $U$ 
    is fixed to zero or not within the multi-parameter fit.
    For our calculations, we employ the one-loop contribution coming 
    from a generic $N$-Higgs doublet model obtained in Ref.~\cite{Grimus:2007if},
    for further details see Appendix~\ref{sec:appRho}. Our analysis is consistent
    with Ref.~\cite{Hernandez:2015rfa} where the interplay between the maximum 
    number of $N$-Higgs doublets and their allowed masses in the oblique parameters
    is discussed.
    
    \item Charged Higgs masses above the lower bound~\cite{Abbiendi:2013hk}: 
    \begin{equation}
         80 \text{ GeV} \lesssim M^\pm_k \;.
    \end{equation}
    
    \item Recently, a search for a Higgs-like particle, $\phi$, 
    decaying into a pair of bottom quarks
    with at least one additional bottom in proton-proton 
    collision was reported~\cite{Aaltonen:2019ied}.
    The following mass range was excluded with 95\% confidence level: 
    \begin{equation}
        100 \text{ GeV} < m_\phi < 300 \text{ GeV} \;.
    \end{equation}
    While not directly comparable with our scenario, our benchmark points nevertheless obey this constraint. 
\end{itemize}

Our four benchmark scenarios are defined by the numerical values of
the Higgs potential parameters as given in 
Table~\ref{tab:NumExam}. The output described by each numerical set is
shown in Table~\ref{tab:NumExam2}.
The first benchmark scenario (Aligned) predicts 
SM-like couplings for all the fermions to the lightest neutral scalar
with mass $m_{h_0} \simeq 125$ GeV. 
The second, third, and fourth benchmark scenarios (Hypo, Hyper, and Mixed) 
feature more drastic departures from the SM-values of Higgs couplings, 
with all couplings being far from the SM expectation.

\section{\label{sec:concl}Conclusions}
\noindent 
Within the SM the huge hierarchy of Yukawa couplings remains a puzzle. In this regard we used the fact that 
the observed fermion masses indicate that the following sets have similar Yukawa couplings: 
$\{ m_t\}$, $\{ m_b,m_\tau,m_c \}$, $\{ m_\mu,m_s \}$ and $\{ m_d,m_u,m_e \}$. 
We have shown that a 4HDM can be constructed that explains this feature. Each set of fermions has its own Higgs doublet. Their vevs are hierarchical which explains the mass hierarchy of the sets. In the model a flavour symmetry was introduced to generate rank $1$ Yukawa matrices. Soft breaking was included in the potential, which makes it possible to induce the smaller vevs by the larger ones, where each smaller vev corresponds to a different broken symmetry, and is thus protected by it. 
All Yukawa couplings take on "natural" values of order $1$. 
In the model neutrino masses are generated via a type-I seesaw mechanism with a Dirac neutrino mass matrix of order of the down-quark mass scale, hence the right-handed singlet Majorana neutrinos are of PeV-scale. 
We have demonstrated that fermions of a given set couple to the SM-like Higgs with the same modified factor. 
In this regard, the clearest signal for this kind
of models is to investigate their coupling to the SM-like Higgs and determine if they are grouped. 
The top quark couples to the SM-like Higgs essentially with the same strength as in the SM.
Benchmark scenarios with definite predictions for those couplings as well as for scalar masses were provided.

Multi-Higgs doublet models face of course problems with 
FCNC. By  singularly aligning the Yukawa matrices we have shown explicitly that those can be evaded. 
This alignment assumes that the Yukawa matrices are related  to the rank $1$ matrices that appear in 
the singular value decomposition of the mass matrices. In this manner, it is in general, 
not only in our model, possible to avoid FCNC while 
simultaneously coupling several Higgs doublets to an individual given fermion. 
Moreover, its equivalence to the most general Yukawa alignment also allows us to state that its
misalignment at the one-loop level is sufficiently small and is consistent with all known phenomenological
tests.
 
The model as well as aspects of singular alignment allow for several follow-up studies 
regarding both model building and  phenomenology.

\appendix
\section{\label{sec:appA}Proof of FCNC Disappearance in the singular basis}
\noindent
Consider a given fermion type coupled to $N$ different
scalar doublets. Its mass matrix would be given by
\begin{align}\label{eq:app1}
	{\bf M} =  v_1 {\bf Y}_1 + 
	v_2 {\bf Y}_2 + \cdots +
	v_N {\bf Y}_N\;,
\end{align} 
where we have assumed that each scalar doublet acquires a
vev. On the other hand, the SVD of the
mass matrix is 
\begin{align}
	{\bf M} = {\bf L}^\dagger \text{diag}(m_1,m_2,m_3) {\bf R} \;,
\end{align}
where ${\bf L}$ and ${\bf R}$ are unitary transformations acting
independently on the left- and right-handed fields.

Using Dirac notation, the SVD can be rewritten as 
\begin{align} \label{eq:appSVD}
	{\bf M} = \sum_i m_i |\ell_i\rangle \langle r_i|\;.
\end{align}
Singular alignment requires assuming 
each Yukawa matrix to be related to the 
rank $1$ matrices, $|\ell_i\rangle \langle r_i|$, 
of the SVD. In general,
we can express the Yukawa matrices as a linear combination
of the rank one singular matrices
\begin{align}
    {\bf Y}_i = \eta_i |\ell_1\rangle \langle r_1|
    + \Omega_i |\ell_2\rangle \langle r_2|
    + \Lambda_i |\ell_3\rangle \langle r_3| \;,
\end{align}
where the parameters $\{\eta,\Omega,\Lambda\}$ are real.

In the mass basis, each Yukawa matrix would take the form,
\begin{align}
    {\bf L} {\bf Y}_i {\bf R}^\dagger =
    \begin{pmatrix}
    \eta_i & 0 & 0\\
    0 & \Omega_i & 0 \\
    0 & 0 & \Lambda_i
    \end{pmatrix} \; .
\end{align}
Therefore, through singularly aligning we have avoided the appearance
of dangerous tree-level FCNC. 

For last, notice that after substitution of the previous relation in  Eq.~\eqref{eq:app1} we obtain 
\begin{align}
m_1 = \sum_j v_j \eta_j \;,\quad
m_2 = \sum_j v_j \Omega_j\;,\quad
m_3 = \sum_j v_j \Lambda_j \;.
\end{align} 

\section{\label{sec:appMixing}Numerical Example for Quark Mixing}
\noindent 
The following singular matrices allow us to reproduce 
exactly the observed mixing in the quark sector
as recently reported in the PDG 2018~\cite{Tanabashi:2018oca}: 
\begin{align}
\begin{split}
    {\bf \Delta}^u_1 & = 
    \begin{pmatrix}
        0.117482 & 0 & 0 \\
        0.984047 e^{+0.485 i} & 0 & 0 \\
        0.133604 e^{+1.27973 i} & 0 & 0
    \end{pmatrix} \;, \quad
    {\bf \Delta}^u_2 = 
    \begin{pmatrix}
        0& 0.0236066 & 0 \\
        0& 0.135386 e^{-3.89058 i} & 0  \\
        0& 0.990512 e^{+0.0259661 i} & 0 
    \end{pmatrix} \;, \\
    {\bf \Delta}^u_3 &= 
    \begin{pmatrix}
        0& 0& 0.992794e^{-0.511327 i}  \\
        0& 0& 0.115423 e^{i \pi}  \\
        0& 0& 0.0321998  e^{i \pi}
    \end{pmatrix} \;, \quad
    {\bf \Delta}^d_1  = 
    \begin{pmatrix}
        0.109076 & 0 & 0 \\
        0.989786 e^{+3.00176 i} & 0 & 0 \\
        0.0917962 e^{+3.52235 i} & 0 & 0
    \end{pmatrix} \;, \\
    {\bf \Delta}^d_2 & = 
    \begin{pmatrix}
        0& 0.0365871 & 0 \\
        0& 0.0886041 e^{+2.59211 i} & 0  \\
        0& 0.995395 e^{+6.27169 i} & 0 
    \end{pmatrix} \;, \quad
    {\bf \Delta}^d_3 = 
    \begin{pmatrix}
        0& 0& 0.99336 e^{-6.13171 i}  \\
        0& 0& 0.111685   \\
        0& 0& 0.0276181 e^{i \pi}  
    \end{pmatrix} ,
\end{split}
\end{align}
where the implied mixing matrix is
\begin{align}
    |{\bf V}^\text{th}_\text{CKM}| = 
    \begin{pmatrix}
    0.97445 & 0.22458 & 0.00364 \\
    0.22442 & 0.97358 & 0.04217 \\
    0.00897 & 0.04137 & 0.999104
    \end{pmatrix} ,
\end{align}
with a Jarlskog invariant of
\begin{align}
    J_q^\text{th} = 3.18 \times 10^{-5} \;.
\end{align}

\section{\label{sec:appConditions}Scalar Mass Matrices}
\noindent 
In this section we discuss the scalar mass matrices. With four Higgs doublets 
there are 4 physical CP-even scalars, 3 pseudoscalars and 3 pairs of charged Higgses. 
Computation of the scalar mass matrices in the limit 
\begin{equation}
   \{ v_t^2,\mu^2_{b},\mu^2_\mu,\mu_d^2\} \gg 
   \{ \mu^2_{tb}, \mu^2_{b\mu}, \mu^2_{\mu d}, v^2_{b}, v^2_{\mu}, v^2_{d}\} \;
\end{equation}
leads to 
\begin{small}
\begin{align}
    {\bf M}_\text{$CP$-even}^2 \simeq 
    \begin{pmatrix}
    2 v_t^2 \lambda_t & 
    2 v_t v_b (XYZ)_{tb} + \mu_{tb}^2 & 
    2v_t v_\mu (XYZ)_{t\mu} &
    2 v_t v_d (XYZ)_{td} \\
    2 v_t v_b (XYZ)_{tb} + \mu_{tb}^2 &
    v_t^2 (XYZ)_{tb} + \mu_b^2 & 
    2 v_b v_\mu (XYZ)_{b\mu} + \mu_{b\mu}^2 &
    2 v_b v_d (XYZ)_{bd} \\
    2 v_t v_\mu (XYZ)_{t\mu} & 
    2 v_b v_\mu (XYZ)_{b\mu} + \mu_{b\mu}^2 &
    v_t^2 (XYZ)_{t\mu} +\mu_\mu^2  & 
    2 v_d v_\mu (XYZ)_{\mu d} + \mu_{\mu d}^2 \\
    2 v_t v_d (XYZ)_{td} & 
    2 v_b v_d (XYZ)_{bd} & 
    2v_\mu v_d (XYZ)_{\mu d} +\mu^2_{\mu d} &
   v_t^2 (XYZ)_{td} +\mu_d^2 
    \end{pmatrix} ,
\end{align}
\end{small}
\begin{align}
    {\bf M}_\text{$CP$-odd}^2 \simeq 
  \begin{pmatrix}
      v_t^2 (X+Y-Z)_{tb} + \mu_b^2 & 0 & 0 \\
     0 & v_t^2 (X+Y-Z)_{t\mu} + \mu_\mu^2 & 0 \\
     0 & 0 & v_t^2 (X+Y-Z)_{td} + \mu^2_d 
   \end{pmatrix}  ,
\end{align}
\begin{align}
    {\bf M}_\text{charged}^2 \simeq 
    \begin{pmatrix}
     v_t^2 X_{tb} + \mu_b^2 & 0 & 0 \\
     0 & v_t^2 X_{t\mu} + \mu_\mu^2 & 0 \\
     0 & 0 & v_t^2 X_{td} + \mu^2_d 
    \end{pmatrix} .
\end{align}
Three necessary but not sufficient conditions may help us 
produce positive definite squared mass sub-matrices.
Given an Hermitian matrix ${\cal A}$ 
\begin{enumerate}
    \item[(i)] all diagonal elements should be positive:  
    \begin{equation}
        [{\cal A}]_{ii} > 0 \;.
    \end{equation}
    \item[(ii)] the sum of any pair of diagonal entries should satisfy 
    \begin{equation}
        [{\cal A}]_{ii} + [{\cal A}]_{jj} > 2 |\Re ([{\cal A}]_{ij})| \;.
    \end{equation}
    \item[(iii)] the largest element should lie on the diagonal.
\end{enumerate}
By virtue of those conditions one may easily derive from the charged scalar matrix that: 
\begin{align}
    X_{tb} > -\frac{\mu^2_b}{v_t^2} \;, \qquad X_{t\mu} > -\frac{\mu^2_\mu}{v_t^2} \;, \qquad
    X_{td} > -\frac{\mu^2_d}{v_t^2} \;.
\end{align}
For the pseudo-scalar matrix: 
\begin{align}
    (X+Y-Z)_{ta} > -\frac{\mu^2_a}{v_t^2} \;, \quad (a=b,\mu,d) \;.
\end{align}
We have not neglected here the off-diagonal contributions to the $CP$-even scalar matrix, as even though
they are very small, they can still influence the Higgs-fermion couplings as already previously discussed.

\section{\label{sec:appRho}Constraints from the $\rho$ parameter}
\begin{figure}[t]
\begin{center}
		\includegraphics[scale=0.6]{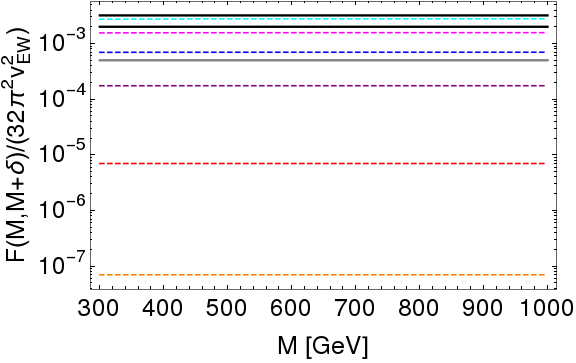}
	\caption{Plot showing the different values for the 
	function $F(x,x+\delta)/(32\pi^2 v^2_\text{EW})$, where
	$\delta=\{1,10,50,100,150,200\} \text{ GeV}$ corresponding, from bottom
	to top, to the dashed lines (orange, red, purple, blue, magenta, cyan). The three continuous
	lines (one gray and two black) correspond to $\Delta \rho = 0.0005$ and its $3\sigma$ upper 
	bound depending on whether the oblique parameter $U$ is fixed to zero or not. 
	The value is strongly dominated by the mass difference, $\delta$, whenever
	scalar masses are in the range $M>300 \text{ GeV}$. }
	\label{fig:rho}
\end{center}
\end{figure}

\noindent 
The one-loop level contribution to the $\rho$ parameter from a theory with $N$ Higgs
doublets has been calculated in Ref.~\cite{Grimus:2007if} and is expressed as
\begin{align}
\begin{split}
    \Delta\rho = \frac{1}{32\pi^2 v_\text{EW}^2}            
    \left[ \sum_{i=2}^N \sum_{j=2}^{2N} |({\cal O}^\dagger {\cal S})_{ij}|^2 
    F\left( M_i^\pm , m_{j}^0 \right)
    - \sum_{i=2}^{2N-1} \sum_{j=i+1}^{2N} |({\cal S}^\dagger {\cal S})_{ij}|^2 
    F\left( m_i^0 , m_j^0 \right) \right. \\ 
    +3 \sum^{2N}_{i=2} |({\cal S}^\dagger {\cal S})_{1i}|^2 
    \left( F\left( M_Z , m_i^0 \right) - F\left( M_W , m_i^0 \right)\right)  
    -3 (F\left( M_Z , m_{h0}^0 \right) - F\left( M_W , m_{h0}^0 \right))
    \Bigg] ,
\end{split}
\end{align}
where 
\begin{equation}
    F(x,y) \equiv 
    \begin{cases}
    \frac{x^2+y^2}{2}-\frac{x^2 y^2}{x^2-y^2}\text{ln}\frac{x^2}{y^2} \;, \qquad \; x\neq y \\
    0 \;, \qquad \qquad \qquad \qquad \qquad x=y
    \end{cases}
\end{equation}
and the matrices ${\cal O}$ and ${\cal S} = {\cal R} \oplus {\cal R}'$ 
are the orthogonal matrices responsible for 
diagonalizing the mass matrices for the charged, $CP$-even and -odd  scalars, respectively.
The function $F(x,y)$ is a positive function, symmetrical under the interchange of its arguments,
and vanishing if and only if the arguments are equal. The behavior of this function has
an interesting property, as it grows linearly with max$(x,y)$, that is, quadratically
with the heaviest-scalar mass, when that mass becomes very large. 
As long as the difference in the scalar masses is small, $\delta \lesssim 200 \text{ GeV}$, 
the maximum value of this function lies within the $3\sigma$ deviation in $\Delta \rho$,
as shown in Fig.~\ref{fig:rho}, even if the masses become very heavy, $M > 300$ GeV. 
This can be seen from Taylor expanding the function, $\delta \ll x$,
\begin{align}
    F(x,x+\delta) = \delta^2 \left[ \frac{2}{3} -\frac{\delta^2}{30x^2}\right] + {\cal O}(\delta^5)\;.
\end{align}
Moreover, realize that these contributions will get further
suppressed by the factors coming from the off-diagonal matrix elements in the product of the orthogonal matrices. 

In the limit in which we are working, mass matrices can be considered to a very good degree
of accuracy to be diagonal, therefore, the maximum amount of contributions in this model will take
the form 
\begin{align}
    \Delta\rho  \simeq \frac{1}{32\pi^2 v_\text{EW}^2}            
    \sum_{i=1}^3  \left[ 
    F\left( M_i^\pm , m_{h,i}^0 \right) +
     F\left( M_i^\pm , m_{A,i}^0 \right) \right] 
      +  0.00017\;.
\end{align}

\acknowledgments
The authors acknowledge useful conversations with Giorgio Busoni, Florian Goertz, Karla Tame-Narvaez, Andreas Trautner  and Stefan Vogl.
The work of WR is supported by the DFG with grant RO 2516/7-1 in the Heisenberg program.
US acknowledges support from CONACYT-M\'exico.

\bibliographystyle{JHEP}	
\bibliography{ulises}

\providecommand{\href}[2]{#2}\begingroup\raggedright\begin{thebibliography}{10}

\bibitem{Weinberg:1977hb}
S.~Weinberg, \emph{{The Problem of Mass}},
  \href{https://doi.org/10.1111/j.2164-0947.1977.tb02958.x}{\emph{Trans. New
  York Acad. Sci.} {\bfseries 38} (1977) 185}.

\bibitem{Porto:2007ed}
R.~A. Porto and A.~Zee, \emph{{The Private Higgs}},
  \href{https://doi.org/10.1016/j.physletb.2008.08.001}{\emph{Phys. Lett.}
  {\bfseries B666} (2008) 491}
  [\href{https://arxiv.org/abs/0712.0448}{{\ttfamily 0712.0448}}].

\bibitem{Porto:2008hb}
R.~A. Porto and A.~Zee, \emph{{Neutrino Mixing and the Private Higgs}},
  \href{https://doi.org/10.1103/PhysRevD.79.013003}{\emph{Phys. Rev.}
  {\bfseries D79} (2009) 013003}
  [\href{https://arxiv.org/abs/0807.0612}{{\ttfamily 0807.0612}}].

\bibitem{Camargo-Molina:2017klw}
J.~E. Camargo-Molina, T.~Mandal, R.~Pasechnik and J.~Wess\'en, \emph{{Heavy
  charged scalars from $c\bar{s}$ fusion: A generic search strategy applied to
  a 3HDM with $\mathrm{U}(1) \times \mathrm{U}(1)$ family symmetry}},
  \href{https://doi.org/10.1007/JHEP03(2018)024}{\emph{JHEP} {\bfseries 18}
  (2018) 024} [\href{https://arxiv.org/abs/1711.03551}{{\ttfamily
  1711.03551}}].

\bibitem{Diaz-Cruz:2019emo}
J.~L. D\'iaz-Cruz, B.~O. Larios-Lopez and M.~A. Perez-de Leon, \emph{{A Private
  SUSY 4HDM with FCNC in the Up-sector}},
  \href{https://arxiv.org/abs/1901.01304}{{\ttfamily 1901.01304}}.

\bibitem{Hill}
C.~T. Hill, P.~A.~N. Machado, A.~E. Thomsen and J.~Turner, \emph{{Scalar
  Democracy}},  \href{https://arxiv.org/abs/1902.07214}{{\ttfamily
  1902.07214}}.

\bibitem{Wang:2006jy}
F.~Wang, W.~Wang and J.~M. Yang, \emph{{Split two-Higgs-doublet model and
  neutrino condensation}},
  \href{https://doi.org/10.1209/epl/i2006-10293-3}{\emph{Europhys. Lett.}
  {\bfseries 76} (2006) 388}
  [\href{https://arxiv.org/abs/hep-ph/0601018}{{\ttfamily hep-ph/0601018}}].

\bibitem{Campos:2017dgc}
M.~D. Campos, D.~Cogollo, M.~Lindner, T.~Melo, F.~S. Queiroz and W.~Rodejohann,
  \emph{{Neutrino Masses and Absence of Flavor Changing Interactions in the
  2HDM from Gauge Principles}},
  \href{https://doi.org/10.1007/JHEP08(2017)092}{\emph{JHEP} {\bfseries 08}
  (2017) 092} [\href{https://arxiv.org/abs/1705.05388}{{\ttfamily
  1705.05388}}].

\bibitem{Cheng:1987rs}
T.~P. Cheng and M.~Sher, \emph{{Mass Matrix Ansatz and Flavor Nonconservation
  in Models with Multiple Higgs Doublets}},
  \href{https://doi.org/10.1103/PhysRevD.35.3484}{\emph{Phys. Rev.} {\bfseries
  D35} (1987) 3484}.

\bibitem{Paschos:1976ay}
E.~A. Paschos, \emph{{Diagonal Neutral Currents}},
  \href{https://doi.org/10.1103/PhysRevD.15.1966}{\emph{Phys. Rev.} {\bfseries
  D15} (1977) 1966}.

\bibitem{Glashow:1976nt}
S.~L. Glashow and S.~Weinberg, \emph{{Natural Conservation Laws for Neutral
  Currents}}, \href{https://doi.org/10.1103/PhysRevD.15.1958}{\emph{Phys. Rev.}
  {\bfseries D15} (1977) 1958}.

\bibitem{Pich:2009sp}
A.~Pich and P.~Tuzon, \emph{{Yukawa Alignment in the Two-Higgs-Doublet Model}},
  \href{https://doi.org/10.1103/PhysRevD.80.091702}{\emph{Phys. Rev.}
  {\bfseries D80} (2009) 091702}
  [\href{https://arxiv.org/abs/0908.1554}{{\ttfamily 0908.1554}}].

\bibitem{Penuelas:2017ikk}
A.~Pe\~nuelas and A.~Pich, \emph{{Flavour alignment in multi-Higgs-doublet
  models}}, \href{https://doi.org/10.1007/JHEP12(2017)084}{\emph{JHEP}
  {\bfseries 12} (2017) 084}
  [\href{https://arxiv.org/abs/1710.02040}{{\ttfamily 1710.02040}}].

\bibitem{Ivanov:2011ae}
I.~P. Ivanov, V.~Keus and E.~Vdovin, \emph{{Abelian symmetries in
  multi-Higgs-doublet models}},
  \href{https://doi.org/10.1088/1751-8113/45/21/215201}{\emph{J. Phys.}
  {\bfseries A45} (2012) 215201}
  [\href{https://arxiv.org/abs/1112.1660}{{\ttfamily 1112.1660}}].

\bibitem{Kagan:2009bn}
A.~L. Kagan, G.~Perez, T.~Volansky and J.~Zupan, \emph{{General Minimal Flavor
  Violation}}, \href{https://doi.org/10.1103/PhysRevD.80.076002}{\emph{Phys.
  Rev.} {\bfseries D80} (2009) 076002}
  [\href{https://arxiv.org/abs/0903.1794}{{\ttfamily 0903.1794}}].

\bibitem{Buras:2010mh}
A.~J. Buras, M.~V. Carlucci, S.~Gori and G.~Isidori, \emph{{Higgs-mediated
  FCNCs: Natural Flavour Conservation vs. Minimal Flavour Violation}},
  \href{https://doi.org/10.1007/JHEP10(2010)009}{\emph{JHEP} {\bfseries 10}
  (2010) 009} [\href{https://arxiv.org/abs/1005.5310}{{\ttfamily 1005.5310}}].

\bibitem{DAmbrosio:2002vsn}
G.~D'Ambrosio, G.~F. Giudice, G.~Isidori and A.~Strumia, \emph{{Minimal flavor
  violation: An Effective field theory approach}},
  \href{https://doi.org/10.1016/S0550-3213(02)00836-2}{\emph{Nucl. Phys.}
  {\bfseries B645} (2002) 155}
  [\href{https://arxiv.org/abs/hep-ph/0207036}{{\ttfamily hep-ph/0207036}}].

\bibitem{Hall:1990ac}
L.~J. Hall and L.~Randall, \emph{{Weak scale effective supersymmetry}},
  \href{https://doi.org/10.1103/PhysRevLett.65.2939}{\emph{Phys. Rev. Lett.}
  {\bfseries 65} (1990) 2939}.

\bibitem{Chivukula:1987py}
R.~S. Chivukula and H.~Georgi, \emph{{Composite Technicolor Standard Model}},
  \href{https://doi.org/10.1016/0370-2693(87)90713-1}{\emph{Phys. Lett.}
  {\bfseries B188} (1987) 99}.

\bibitem{Buras:2000dm}
A.~J. Buras, P.~Gambino, M.~Gorbahn, S.~Jager and L.~Silvestrini,
  \emph{{Universal unitarity triangle and physics beyond the standard model}},
  \href{https://doi.org/10.1016/S0370-2693(01)00061-2}{\emph{Phys. Lett.}
  {\bfseries B500} (2001) 161}
  [\href{https://arxiv.org/abs/hep-ph/0007085}{{\ttfamily hep-ph/0007085}}].

\bibitem{Ilakovac:1994kj}
A.~Ilakovac and A.~Pilaftsis, \emph{{Flavor violating charged lepton decays in
  seesaw-type models}},
  \href{https://doi.org/10.1016/0550-3213(94)00567-X}{\emph{Nucl. Phys.}
  {\bfseries B437} (1995) 491}
  [\href{https://arxiv.org/abs/hep-ph/9403398}{{\ttfamily hep-ph/9403398}}].

\bibitem{TheMEG:2016wtm}
{\scshape MEG} collaboration, \emph{{Search for the lepton flavour violating
  decay $\mu ^+ \rightarrow \mathrm {e}^+ \gamma $ with the full dataset of the
  MEG experiment}},
  \href{https://doi.org/10.1140/epjc/s10052-016-4271-x}{\emph{Eur. Phys. J.}
  {\bfseries C76} (2016) 434}
  [\href{https://arxiv.org/abs/1605.05081}{{\ttfamily 1605.05081}}].

\bibitem{Bellgardt:1987du}
{\scshape SINDRUM} collaboration, \emph{{Search for the Decay mu+ ---> e+ e+
  e-}}, \href{https://doi.org/10.1016/0550-3213(88)90462-2}{\emph{Nucl. Phys.}
  {\bfseries B299} (1988) 1}.

\bibitem{Aaij:2014azz}
{\scshape LHCb} collaboration, \emph{{Search for the lepton flavour violating
  decay $\tau^{-}\rightarrow \mu^{-} \mu^{+}\mu^{-}$}},
  \href{https://doi.org/10.1007/JHEP02(2015)121}{\emph{JHEP} {\bfseries 02}
  (2015) 121} [\href{https://arxiv.org/abs/1409.8548}{{\ttfamily 1409.8548}}].

\bibitem{Aubert:2009ag}
{\scshape BaBar} collaboration, \emph{{Searches for Lepton Flavor Violation in
  the Decays tau+- ---> e+- gamma and tau+- ---> mu+- gamma}},
  \href{https://doi.org/10.1103/PhysRevLett.104.021802}{\emph{Phys. Rev. Lett.}
  {\bfseries 104} (2010) 021802}
  [\href{https://arxiv.org/abs/0908.2381}{{\ttfamily 0908.2381}}].

\bibitem{Hambye:2013jsa}
T.~Hambye, \emph{{CLFV and the origin of neutrino masses}},
  \href{https://doi.org/10.1016/j.nuclphysbps.2014.02.004}{\emph{Nucl. Phys.
  Proc. Suppl.} {\bfseries 248-250} (2014) 13}
  [\href{https://arxiv.org/abs/1312.5214}{{\ttfamily 1312.5214}}].

\bibitem{Cirigliano:2005ck}
V.~Cirigliano, B.~Grinstein, G.~Isidori and M.~B. Wise, \emph{{Minimal flavor
  violation in the lepton sector}},
  \href{https://doi.org/10.1016/j.nuclphysb.2005.08.037}{\emph{Nucl. Phys.}
  {\bfseries B728} (2005) 121}
  [\href{https://arxiv.org/abs/hep-ph/0507001}{{\ttfamily hep-ph/0507001}}].

\bibitem{Cirigliano:2006su}
V.~Cirigliano and B.~Grinstein, \emph{{Phenomenology of minimal lepton flavor
  violation}},
  \href{https://doi.org/10.1016/j.nuclphysb.2006.06.021}{\emph{Nucl. Phys.}
  {\bfseries B752} (2006) 18}
  [\href{https://arxiv.org/abs/hep-ph/0601111}{{\ttfamily hep-ph/0601111}}].

\bibitem{Sirunyan:2018koj}
{\scshape CMS} collaboration, \emph{{Combined measurements of Higgs boson
  couplings in proton-proton collisions at $\sqrt{s}=$ 13 TeV}},
  {\emph{Submitted to: Eur. Phys. J.} (2018) }
  [\href{https://arxiv.org/abs/1809.10733}{{\ttfamily 1809.10733}}].

\bibitem{Wigner:1946zz}
E.~P. Wigner, \emph{{Resonance Reactions and Anomalous Scattering}},
  \href{https://doi.org/10.1103/PhysRev.70.15}{\emph{Phys. Rev.} {\bfseries 70}
  (1946) 15}.

\bibitem{Wigner:1947zz}
E.~P. Wigner and L.~Eisenbud, \emph{{Higher Angular Momenta and Long Range
  Interaction in Resonance Reactions}},
  \href{https://doi.org/10.1103/PhysRev.72.29}{\emph{Phys. Rev.} {\bfseries 72}
  (1947) 29}.

\bibitem{Bell:2016obu}
N.~Bell, G.~Busoni, A.~Kobakhidze, D.~M. Long and M.~A. Schmidt,
  \emph{{Unitarisation of EFT Amplitudes for Dark Matter Searches at the LHC}},
  \href{https://doi.org/10.1007/JHEP08(2016)125}{\emph{JHEP} {\bfseries 08}
  (2016) 125} [\href{https://arxiv.org/abs/1606.02722}{{\ttfamily
  1606.02722}}].

\bibitem{Bell:2016ekl}
N.~F. Bell, G.~Busoni and I.~W. Sanderson, \emph{{Self-consistent Dark Matter
  Simplified Models with an s-channel scalar mediator}},
  \href{https://doi.org/10.1088/1475-7516/2017/03/015}{\emph{JCAP} {\bfseries
  1703} (2017) 015} [\href{https://arxiv.org/abs/1612.03475}{{\ttfamily
  1612.03475}}].

\bibitem{Tanabashi:2018oca}
{\scshape Particle Data Group} collaboration, \emph{{Review of Particle
  Physics}}, \href{https://doi.org/10.1103/PhysRevD.98.030001}{\emph{Phys.
  Rev.} {\bfseries D98} (2018) 030001}.

\bibitem{Grimus:2007if}
W.~Grimus, L.~Lavoura, O.~M. Ogreid and P.~Osland, \emph{{A Precision
  constraint on multi-Higgs-doublet models}},
  \href{https://doi.org/10.1088/0954-3899/35/7/075001}{\emph{J. Phys.}
  {\bfseries G35} (2008) 075001}
  [\href{https://arxiv.org/abs/0711.4022}{{\ttfamily 0711.4022}}].

\bibitem{Hernandez:2015rfa}
A.~E. C\'arcamo~Hern\'andez, S.~Kovalenko and I.~Schmidt, \emph{{Precision
  measurements constraints on the number of Higgs doublets}},
  \href{https://doi.org/10.1103/PhysRevD.91.095014}{\emph{Phys. Rev.}
  {\bfseries D91} (2015) 095014}
  [\href{https://arxiv.org/abs/1503.03026}{{\ttfamily 1503.03026}}].

\bibitem{Abbiendi:2013hk}
{\scshape ALEPH, DELPHI, L3, OPAL, LEP} collaboration, \emph{{Search for
  Charged Higgs bosons: Combined Results Using LEP Data}},
  \href{https://doi.org/10.1140/epjc/s10052-013-2463-1}{\emph{Eur. Phys. J.}
  {\bfseries C73} (2013) 2463}
  [\href{https://arxiv.org/abs/1301.6065}{{\ttfamily 1301.6065}}].

\bibitem{Aaltonen:2019ied}
{\scshape CDF} collaboration, \emph{{Search for Higgs-like particles produced
  in association with bottom quarks in proton-antiproton collisions}},
  {\emph{Submitted to: Phys. Rev. D} (2019) }
  [\href{https://arxiv.org/abs/1902.04683}{{\ttfamily 1902.04683}}].

\end{thebibliography}\endgroup

\end{document}